\documentclass
[
 amsmath,amssymb,
 aps,
prb, 10pt
]{revtex4-2}
\usepackage{xcolor}
\usepackage{CJK}

\usepackage{graphicx}
\usepackage{dcolumn}
\usepackage{bm}

\begin{document}

\preprint{APS/123-QED}
\begin{CJK}{UTF8}{gbsn} 
\title{Characterizing phase transitions and criticality in non-Hermitian extensions of the XY model}

\author{D. C. Liu (刘东昌)}
 \email{dongchang.liu@anu.edu.au}
 \author{Murray T. Batchelor}
 \email{murray.batchelor@anu.edu.au}
\affiliation{
 Mathematical Sciences Institute, Australian National University, ACT 2601, Canberra, Australia
}%

\date{\today}

\begin{abstract}
In this work we study non-Hermitian extensions of the paradigmatic spin-1/2 XY chain in a magnetic field. Using the mapping of the model to free fermion form, we provide analytical insights into the energy spectrum of the non-Hermitian model and establish an intrinsic
connection between the quasi-energies and topological invariants. We also use exact diagonalization as a supplementary method to examine the performance of biorthogonal-based expectation values. Our results confirm that the theoretical analysis is consistent with the numerical results, with the extended phase diagram determined via the analytical solution and the critical behavior of the fidelity and entanglement. The entanglement transition goes hand in hand with the non-Hermitian topological phase transition. Like the Hermitian case, we analyze the critical behavior using finite-size scaling. Our results show that non-Hermiticity can induce the system into a new universality class with unusual critical exponent. We also emphasize the ability of the Loschmidt echo to characterize potential topological phase transitions and introduce the average of the Loschmidt echo to describe phase transitions in non-Hermitian systems.

\end{abstract}

\maketitle
\end{CJK}

\section{Introduction}

As a fundamental concept in modern physics, quantum phase transitions (QPTs) occur when a sudden change takes place in the ground state of a system driven by external parameters \cite{Sachdev_2011}. Most QPTs can be described within Landau-Ginzburg-Wilson theory and are characterized by an order parameter and energy level-crossing behavior \cite{Ginzburg2009, HOHENBERG20151}. Typically, studies of quantum phase transitions assume the Hamiltonian to be Hermitian. It is both intriguing and instructive however, to investigate quantum phase transitions in non-Hermitian (NH) systems, where the transition is driven by the competition between the real and imaginary parts of a parameter \cite{doi:10.1126/science.abe9869, PhysRevLett.122.237601}. In general, a NH Hamiltonian has complex eigenvalues, in which the real and imaginary parts respectively represent the energy and give the information on the decay rate of the associated eigenstates \cite{PhysRevA.94.053615}. One remarkable type of phase transition in NH many-body systems is induced by certain symmetries, such as invariance under parity-time ($\mathcal{PT}$) symmetry or intrinsic rotation-time-reversal ($\mathcal{RT}$) symmetry. It has been realized that a Hamiltonian satisfying these symmetries can exhibit an entirely real spectrum \cite{PhysRevLett.80.5243, PhysRevA.87.012114}. The NH quantum phase transition can occur between a phase with a real energy spectrum preserved by symmetry and a symmetry-broken phase with a complex energy spectrum. This type of transition typically occurs across the entire energy spectrum. However, recent studies, such as those on the Lipkin-Meshkov-Glick \cite{PhysRevA.95.010103} and Hatano-Nelson \cite{PhysRevB.106.L121102, PhysRevA.109.042208} models, suggest that NH phase transitions can also occur between specific excited states. With the emergence of NH phase transitions, a natural question is if NH phase transitions exhibit unusual critical behavior. Indeed, some existing studies for NH systems clarifies the observation of critical behavior typical of Hermitian systems \cite{PhysRevB.102.245145, PhysRevX.4.041001,Sun2021}. Recent work illustrates  that the critical exponent relies on the discrepancy between symmetry class in the NH Anderson phase transition \cite{PhysRevLett.126.090402, PhysRevResearch.4.L022035}. The formulation and deeper understanding of NH quantum phase transitions remains an area needing further exploration.

On the other hand, topological phase transitions, which are beyond the paradigm of Landau-Ginzburg theory, have also attracted considerable attention, partly owing to the intriguing nature of the phase itself and associated robust edge states, appearing in the study of band structure and generally given by the properties of wave functions \cite{PhysRevLett.61.2015,RevModPhys.88.035005,RevModPhys.88.021004}. In recent years, topological band theory has been further developed in NH systems \cite{PhysRevLett.120.146402}.
To date, 38 classes of non-Hermitian topological phases have been recognized and characterized based on their intrinsic symmetries \cite{PhysRevX.9.041015}, compared to the 10 types of topological phases in Hermitian counterparts. A unique feature of the NH system is the skin effect, representing anomalous localization of an extensive number of eigenstates in the presence of non-Hermiticity, which has no analog in Hermitian systems \cite{PhysRevLett.121.086803, PhysRevX.8.031079, Xiao2020}. Originating from the topological invariant, the nature of the bulk-boundary correspondence is altered due to the extreme sensitivity of the bulk to boundary conditions caused by the skin effect. Additionally, as a consequence of the skin effect, the macroscopic particle flow is confined to the boundaries. It inhibits the growth of entanglement entropy and makes the entropy adhere to the area law \cite{RevModPhys.82.277,PhysRevX.13.021007}. Various types of NH-induced entanglement transitions have been evaluated in gapped or critical NH fermion lattices and NH quantum spin chains \cite{PhysRevB.109.024204, PhysRevB.109.024306,PhysRevB.107.L020403, PhysRevLett.126.170503}. At the same time, the topological states are shown to be intrinsically related to the quantum dynamics \cite{PhysRevB.97.064304}, and this connection has recently been brought to NH topological properties through the critical behavior of the Loschmidt echo \cite{PhysRevA.98.022129,Zhou_2021}. NH quantum dynamics can be realized under continuous observation without quantum jumps in experiments \cite{Daly2014}. 

In this article, we investigate QPTs and criticality in the extended NH XY model from the viewpoint of the quasi-energy building blocks making up the free fermion eigenspectrum. This approach brings a general perspective to recent studies of various NH extensions of the XY and quantum Ising models. 
The phase diagram is determined from NH band theory, combined with the exact solution via free fermions. Moreover, the topological invariant winding number is connected to the quasi-energy degeneracy in NH systems, and the zero value of quasienergy indicates the existence of topology. In the model under consideration, the $\mathcal{PT}$-phase transition occurs in the excited states when the system is subjected to a purely imaginary field. We also estimate the critical exponent $\nu$ from the finite-size scaling behavior of the biorthogonal fidelity, and extract the critical points in the thermodynamic limits, matching with the theoretical derivation. Numerical evidence shows that the interaction between loss and gain in the system does not affect its critical behavior, with critical exponent that of the Ising class.  However, we see that the NH anisotropy interaction introduces an unusual type of universality class when the system undergoes phase transition between two topological phases in which the central charge is unchanged. From the perspective of entanglement entropy, the phase transition in the model is accompanied by an entanglement transition, where the entropy in the topological phase exhibits an area-law scaling. We also confirm the theoretically predicted connection between NH topology and dynamic quantum phase transitions by examining the Loschmidt echo numerically. In particular, the Loschmidt echo is seen to present different oscillation behavior if the quenched Hamiltonian is chosen in different phases, and the time average behavior of Loschmidt echo is introduced to differentiate the phase.

\section{Model and Hamiltonians}

The spin-$\frac{1}{2}$ XYh model is defined by the Hamiltonian 
\begin{equation}\label{eq1}
H= - \sum^{N-1}_{n=1}\left(\frac{1+\gamma}{2}\sigma^{x}_n\sigma^{x}_{n+1}+\frac{1-\gamma}{2}\sigma^{y}_n\sigma^{y}_{n+1} \right) +h \sum^{N}_{n=1} \sigma^{z}_n,
\end{equation}
where $\sigma_n^{\mu}$($\mu=x, y, z$) are the Pauli operators. The Hamiltonian is defined for open boundary conditions (OBC), with the boundary terms $\sigma _{N+1}^\mu=\sigma_1^\mu$ added when periodic boundary conditions (PBC) are considered. This model is recognized as a textbook example for exploring magnetic QPTs. The model reduces to the XY model for $h=0$ and the quantum Ising model for $\gamma=1$. In the XYh model, two types of QPTs (anisotropic and Ising phase transitions) can be characterized by different critical behavior \cite{PFEUTY197079,LIEB1961407,McCoy_book}.

Generally, non-Hermiticity can arise from on-site imaginary potentials in tight-binding models \cite{PhysRevA.80.052107} or imaginary magnetic fields in quantum spin models \cite{PhysRevA.90.012103,STARKOV2023169268}. The complex coupling between spins can also reflect non-Hermiticity, with ultra cold atomic experiments providing an ideal platform for the simulation of similarly imaginary interaction \cite{PhysRevLett.126.110404, Sponselee_2019,Buca_2020}. Here we extend the parameters $\gamma$ and $h$ appearing in the XYh Hamiltonian (\ref{eq1}) to the complex plane, defining $\gamma=\gamma_{R}+\mathrm{i} \, \gamma_{I}$ and $h=h_{R}+\mathrm{i} \, h_{I}$. In particular, we focus here on three special cases, each defining a NH variant of the XYh model. These are the quantum Ising model with complex field, defined by the Hamiltonian
\begin{equation}\label{eq2}
H_1= - \sum^{N-1}_{n=1}\sigma^{x}_n\sigma^{x}_{n+1} + h \sum^{N}_{n=1} \sigma^{z}_n.
\end{equation}
The XY model with purely complex field, given by 
\begin{equation}\label{eq3}
H_2=  - \sum^{N-1}_{n=1}\left(\frac{1+\gamma_{R}}{2}\sigma^{x}_n\sigma^{x}_{n+1}+\frac{1-\gamma_{R}}{2}\sigma^{y}_n\sigma^{y}_{n+1} \right) + \, \mathrm{i} \, h_{I} \sum^{N}_{n=1} \sigma^{z}_n \,.
\end{equation}
The NH XY model with complex $\gamma$ and real field defined by 
\begin{equation}\label{eq4}
H_3=  - \sum^{N-1}_{n=1}\left(\frac{1+{\gamma}}{2}\sigma^{x}_n\sigma^{x}_{n+1}+\frac{1-{\gamma}}{2}\sigma^{y}_n\sigma^{y}_{n+1} \right) + \, h_{R}\sum^{N}_{n=1} \sigma^{z}_n \,.
\end{equation}

Hamiltonians $H_1$ and $H_2$ are equivalent to the two special cases of the Kitaev Ising chain with onsite particle losses after mapping to fermions  \cite{Zhou_2021,PhysRevA.94.022119}. One of the cases in $H_3$ has been considered recently by taking $h_{R}=0$, reducing to the $\mathcal{RT}$ symmetric XY model \cite{PhysRevA.87.012114}. Also, the NH Hamiltonian with the competition between the imaginary anisotropic interaction and external field, which can be seen as the NH XY model with $\mathcal{RT}$ symmetry, is not considered here, as it has already been studied in detail very recently \cite{PhysRevA.110.012226, PhysRevB.110.014403}.

\section{Methodology}

For general NH systems with Hamiltonian $H \neq H^{\dag}$, one must consider two distinct bases \cite{doi:10.1080/00018732.2021.1876991,Brody_2014}, obtained from
\begin{equation}\label{eq5}
\begin{aligned}
      & ~ H \, \psi_n^R = E_n \, \psi_n^R \,, \\
      & H^{\dag} \, \psi_n^L = E_n{}^{*} \, \psi_n^L \,,
\end{aligned}
\end{equation}
in which the left eigenstate $\psi^L_n$ and right eigenstate $\psi^R_n$ correspond to the $n$th eigenvalue $E_n $. The left and right eigenstates satisfy the bi-orthonormal relationship
\begin{equation}\label{eq6}
\langle \psi_n^L | \psi_m ^R\rangle=\delta_{nm} ,
\end{equation}
and completeness relationship
\begin{equation}\label{eq7}
\sum_n | \psi_m ^R\rangle \langle \psi_n^L |=1.
\end{equation}
As a result, many more novel properties emerge in NH systems since we have more degrees of freedom in choosing $ | \psi_n^L \rangle$ independent of $ | \psi_n^R \rangle$ compared to the Hermitian case. The physical exceptional value differences appear owing to the flexibility in how to choose the basis mutually \cite{PhysRevResearch.5.033181,Chen_2024}.

\subsection{Free-fermion solution for the open chain}

The general XYh model can be mapped to free (non-interacting) fermions. The basic property of a free-fermionic system is that all energies in the energy spectrum take the form
\begin{equation}\label{eq8}
E=\sum_{j=1}^{N} \pm \, \epsilon_{j},
\end{equation}
where $\epsilon_{j}$ is the $j$th quasi-energy which depends on the system parameters. This simple structure holds exactly for finite system sizes with OBC.   
It yields all $2^N$ energies in the spectrum. In this picture, the groundstate is 
\begin{equation}\label{eq9}
E=- \sum_{i=1}^N \epsilon_{j}. 
\end{equation}

The free-fermion solution can be directly extended to NH cases resulting from turning the system parameters into complex parameters. While the quasi-energies $\epsilon_j$ are ensured to be positive and distinct for the Hermitian case, they become complex for non-Hermitian models. 
As the system parameters are varied, a pair of quasienergies may become equal. Such quasi-energy degeneracies are able to provide insights into NH physics, for example, with the formation of Exceptional Points (EPs), which were studied recently \cite{10.21468/SciPostPhys.15.1.016} in the context of the more general Baxter-Fendley free parafermion model \cite{BAXTER1989155,Fendley_2014,Brief_History} and most recently in the dimerized Hatano-Nelson model with staggered potentials \cite{Sirker_2024}.

In this context, the relative difference between the smallest two quasi-energies, between which a degeneracy appears, is defined as
\begin{equation}\label{eq10}
    \Delta_{12} =\frac{|\epsilon_1-\epsilon_{2}|}{|\epsilon_1+\epsilon_{2}|},
\end{equation}
where $\epsilon_1$ and $\epsilon_2$  are the two smallest quasi-energy in absolute value.

For the NH XYh model, the solution in terms of free fermions can be extended to complex $\gamma$ and $h$. For this model, the quasi-energies follow directly from the eigenvalues of the $N \times N$ matrix (see, e.g., Ref.~\cite{Asakawa_1995})
\begin{equation}\label{eq11}
C = 
\begin{pmatrix}
h^2 + y^2 & h & x \, y & 0 & \cdots & 0 \\
h & h^2 + x^2 + y^2 & h & x \, y & \cdots & 0 \\
x \, y & h & h^2 + x^2 + y^2 & h & \cdots & 0 \\
0 & x \, y & h & h^2 + x^2 + y^2 & \cdots & h \\
\vdots & \vdots & \vdots & \vdots & \ddots & \vdots \\
0 & 0 & 0 & \cdots & h & h^2 + x^2 
\end{pmatrix} \,,
\end{equation}
where $x=\frac{1}{2}(1+\gamma)$ and $y=\frac{1}{2}(1-\gamma)$.
Specifically, $\epsilon_j= a_j ^{{1}/{2}}$ where $a_j$ are the eigenvalues of the matrix $C$.
A test on the validity of the correctness of the free fermion solution for complex couplings is given in Appendix A, which also shows the typical structure of the complex energy spectrum. 


\subsection{Non-Hermitian topological invariants}

The Hamiltonian~(\ref{eq1}) maps onto spinless Fermi operators via Jordan-Wigner transformation~\cite{Jordan1928} resulting in the spinless fermion form 
\begin{equation}\label{eq12}
    H= -  \sum^{N-1}_{1}\left( c_n^{\dag}c_{n+1}+\gamma \, c_n^{\dag}c_{n+1}^{\dag} +h.c. \right) +2h \sum^{N}_{1} \left(c_n^{\dag}c_{n}-\frac{1}{2}\right).
\end{equation}
Here $c_n$ and $c_n^{\dag}$ are the fermionic annihilation and creation operators at site $n$. After applying the Fourier transformation $c_n=\frac{1}{\sqrt{L}}\sum_k e^{\mathrm{i} k n} c_k$ in PBC, the Hamiltonian $H$ can be expressed in the form of momentum space as $H=\sum_k \Psi_k^{\dag} H(k) \Psi_k$ with Nambu spinor operator $\Psi_k^{\dag}=\left( c_k^{\dag}, c_{-k}\right)$ in the quasimomentum $k\in [-\pi,\pi)$. In the theory of non-Hermitian topological phases, if $H(k)$ possesses the chiral symmetry $\mathcal{S} H(k) \mathcal{S}=-H(k)$, we can define the winding number to characterize the topological phases in the system~\cite{PhysRevA.97.052115}. By substituting the form of $\gamma$ and $h$ in the complex plane, $H(k)$ is given by
\begin{equation}\label{eq13}
H(k)=\begin{pmatrix}
\cos(k)-h_{R}-\mathrm{i}\, h_{I}& \mathrm{i} \, \gamma_{R}\sin(k)-\gamma_{I}\sin(k)\\
- \mathrm{i} \, \gamma_{R}\sin(k)+\gamma_{I}\sin(k)& h_{R}+\mathrm{i} \,h_{I}-\cos(k)\\
\end{pmatrix},
\end{equation}
which can be further expressed in terms of Pauli matrices $\sigma_y$ and $\sigma_z$ as   
\begin{equation}\label{eq14}
H(k)=[u_y(k)-\mathrm{i} \, v_y(k)]\sigma_y+[u_z(k)-\mathrm{i} \, v_z(k)]\sigma_z,
\end{equation}
with $u(k)$ and $v(k)$ defined by 
\begin{equation}
\begin{aligned}
    u_y(k) &= - \gamma_R \sin(k), \quad
    u_z(k) = \cos(k) - h_R, \\
    v_y(k) &= \gamma_I \sin(k), \qquad
    v_z(k) = h_I.
\end{aligned}
\end{equation}

The winding number, which describes the accumulated change of winding angle $\phi(k)\equiv\arctan [\frac{u_y(k)-\mathrm{i}  v_y(k)}{u_z(k)-\mathrm{i}  v_z(k)}]$ over the first Brillouin zone (BZ), is defined as 
\begin{equation}\label{eq15}
w=\frac{1}{2 \pi} \int_{-\pi}^{\pi} \mathrm{d}k \, \partial_k \phi(k).
\end{equation}
Previous studies~\cite{PhysRevResearch.2.023043, PhysRevA.97.052115} have already shown that the value of $w$ is always real whereas $\phi(k)$ is in general complex, as the result of the imaginary part of $\phi(k)$ does not wind in the first BZ, which is a real continuous periodic function of $k$.

\subsection{Biorthogonal fidelity}

The fidelity, as the overlap between two states, is widely used in the quantum information sciences as a measurement of the similarity of two quantum states \cite{doi:10.1142/S0217979210056335, PhysRevE.111.014110}. This quantity works as a probe to detect phase transitions in Hermitian many-body systems and has recently been proven to be equally useful in the study of NH systems, for which the definition of fidelity is modified as it is necessary to replace the conventional inner product with a metricized inner product \cite{Tu2023generalpropertiesof,PhysRevA.99.042104,PhysRevResearch.3.013015}. The groundstate fidelity for a NH system $H(\lambda) = H_0 + \lambda H^\prime $ is 
\begin{equation}\label{eq16}
    F=\sqrt{ \langle \psi_0^L (\lambda+ \delta \lambda) | \psi_0^R (\lambda) \rangle \langle \psi_0^L(\lambda) | \psi_0^R (\lambda+ \delta \lambda) \rangle } \,.
\end{equation}
Similar to that for the Hermitian case, the corresponding fidelity susceptibility $\chi_F$ is given by
\begin{equation}\label{eq17}
    \chi_F= \frac{1}{N} \lim_{\delta \lambda \to 0} \frac{-2 \ln F}{\delta \lambda ^2 }.
\end{equation}
As for NH systems, the correlation length critical exponent $\nu$ can be directly extracted from the finite-size scaling behavior of the fidelity susceptibility \cite{Sun2021} via 
\begin{equation}\label{eq18}
\chi_F^{\mathrm{max}}=N^{\frac{2}{\nu} - 1}.
\end{equation}

\subsection{Biorthogonal entropy}

Quantum entanglement plays a crucial role in characterizing and gaining a deep understanding of many-body quantum systems \cite{PhysRevLett.133.100402,SU2022127005}. For a bipartite system  divided into part $A$ and part $B$, the entanglement entropy between $A$ and $B$ is introduced by 
\begin{equation}\label{eq19}
    S_A= - \text{Tr}_A(\rho_A\ln\rho_A),
\end{equation}
where $\rho_A$ is the reduced density matrix of subsystem $A$ by tracing out subsystem $B$. For the $N$-site systems, one has subsystems $A=\{1,\ldots,L_A\}$ and $B=\{L_A+1,\ldots,N\}$, where $L_A$ is the size of subsystem $A$. As the extension of entanglement entropy, a generic entanglement entropy used to measure NH systems can then be defined using the reduced density matrix in terms of the biorthogonal basis~\cite{PhysRevResearch.2.033069,Brody_2014,Chen_2024}. In terms of the biorthogonal basis, the biorthogonal reduced density matrix for the ground state is defined as
\begin{equation}\label{eq20}
        \rho^{RL} = \frac{| \psi_0 ^R\rangle \langle \psi_0 ^L |}{\text{Tr} (| \psi_0 ^R \rangle \langle \psi_0 ^L |)}.
\end{equation}

Meanwhile, the subsystem size scaling of the entanglement entropy is well known to give useful information about conformally invariant systems~\cite{Calabrese_2009}. The scaling behavior can be examined as
\begin{equation}\label{eq21}
    S(L_A) \sim \frac{c}{3} \ln \left(\sin{ \frac{\pi L_A}{N}}\right)+ \mathrm{const.}\,.
\end{equation}
Around the EPs in NH systems, the central
charge has been found to be negative, and argued to be described by nonunitary conformal field theory~\cite{10.21468/SciPostPhys.12.6.194,PhysRevResearch.2.033069,PhysRevB.106.174517,10.21468/SciPostPhys.7.5.069}.

\subsection{Dynamic properties}

We are also interested in the dynamic behavior in NH systems where the time evolution operator is no longer unitary. Recent studies have shown that non-analytical dynamic behavior is considered to be a useful tool in the study of NH systems~\cite{PhysRevA.110.012226, PhysRevResearch.2.023043}, although only the right state of a system can be prepared in experiments~\cite{Hu:24}.
The time evolution is started from the ordered initial state $|\psi(0)\rangle=|\uparrow\uparrow\uparrow...\uparrow\uparrow\uparrow\rangle$ which is, for example, able to be measured in the Rydberg atom~\cite{PhysRevLett.131.080403}. According to the Hamiltonian $H$, the evolved state takes the form
\begin{equation}\label{eq22}
    |\psi(t)\rangle =\frac{e^{-\mathrm{i} Ht}|\psi(0)\rangle}{||e^{-\mathrm{i} Ht}|\psi(0)\rangle||},
\end{equation}
where the normalization factor $||\psi \rangle||=\sqrt{|\langle\psi|\psi\rangle|}$ takes into consideration the fact that the norm is not conserved due to the lack of unitarity~\cite{PhysRevLett.123.090603}. With the normalized state $|\psi(t)\rangle $, the Loschmidt echo 
\begin{equation}\label{eq23}
    \mathcal{L}=\frac{|\langle\psi(0)|\psi(t)\rangle|^2}{\langle\psi(t)|\psi(t)\rangle} \,,
\end{equation}
can be examined as a measurement of dynamical quantum phase transitions
In this context, the Loschmidt echo can be interpreted as the return probability of the initial state during time evolution. The Loschmidt echo measures the ability of quantum evolution to reverse upon an imperfect time reversal~\cite{PhysRevB.100.184313,GORIN200633}. In our ED simulation, the time-evolved operator $e^{-\mathrm{i} Ht}$ governed by $H$ defined as Hamiltonians $H_1$, $H_2$ and $H_3$ is discretized using time steps of $dt = 0.01$, with the time-evolved state $|\psi(t)\rangle$ is subsequently computed using the fourth-order Runge-Kutta method.

\section{Results and discussion}

\begin{figure*}
\centerline{\includegraphics[width=1.2\linewidth]{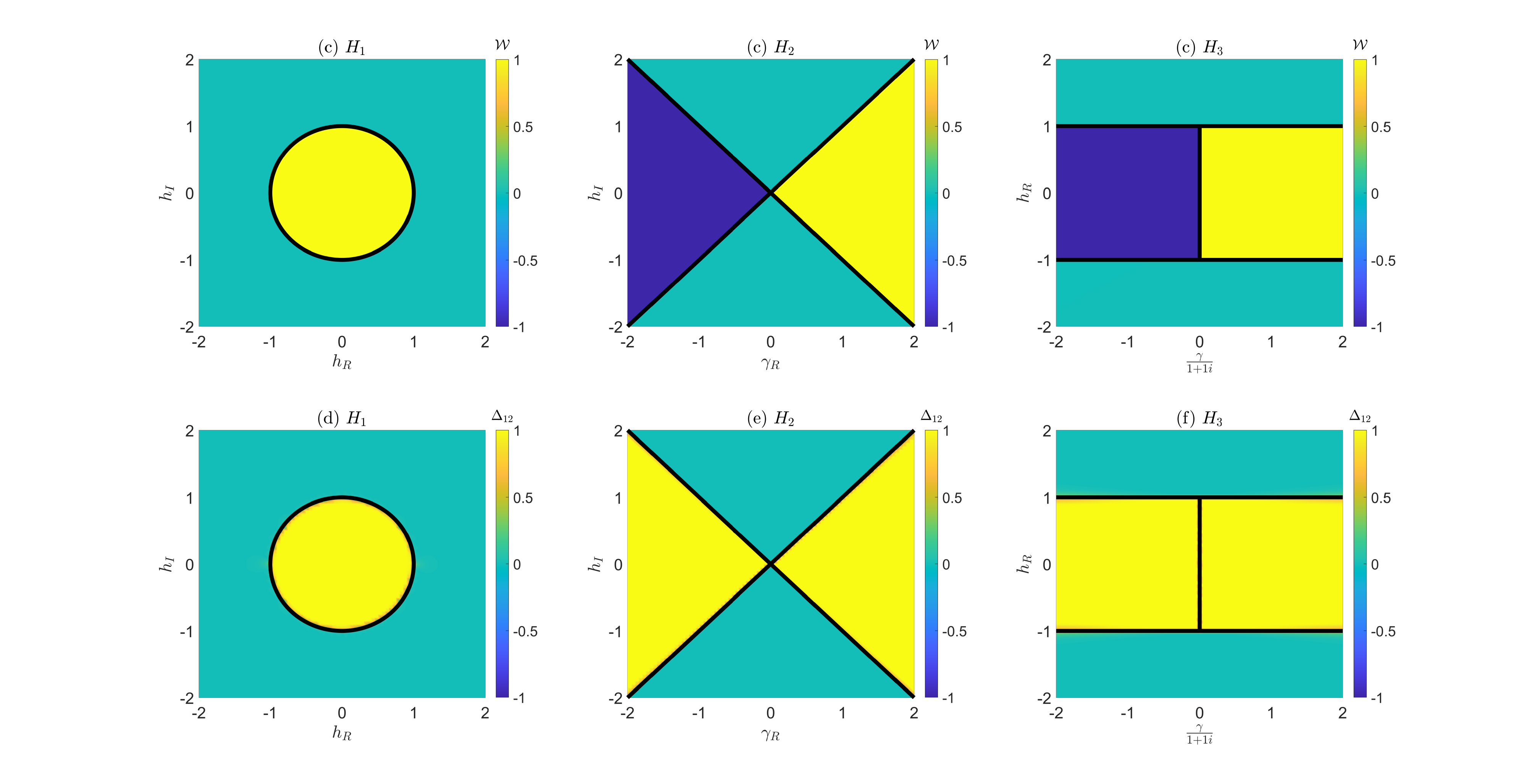}}
\caption{(a), (b), and (c) The phase diagram for the NH XYh model with respect to different model parameters obtained through topological winding number $w$. (d), (e) and (f) are the corresponding phase diagrams obtained from quasi-energy  degeneracy via the free fermion solution with system size $N=300$ and OBC. } 
\label{Fig1}
\end{figure*}

We first present the phase diagram for Hamiltonians $H_1$, $H_2$ and $H_3$ in Fig.~\ref{Fig1} (a)-(c) obtained by using the non-Hermitian winding number $w$. It is seen that the topological phase with $|w|=1$ and trivial phase with $w=0$ are separated by the phase boundary $h_{R}^2+h_{I}^2=1$ for Hamiltonian $H_1$, and the phase boundary located at $\gamma_R=\pm h_{I}$ for Hamiltonian $H_2$. These phase boundaries are consistent with the equation $h_{R}^2+{h_{I}^2}/{\gamma_{R}^2}=1$ obtained by solving the dispersion relation in corresponding momentum space for Hamiltonian $H(k)$ when taking $\gamma_{I}=0$ (see also, e.g.,~Ref.~\cite{Pi_2021}). 
The Hamiltonian $H_3$ considers the contribution of imaginary anisotropic interaction $\gamma$ to the topological properties. In  Fig.~\ref{Fig1} (c), one can see that the introduction of the non-Hermitian anisotropic interaction does not affect the phase diagram, and the phase boundaries are located the same as for the Hermitian XYh model. We confirm this result by calculating the distribution of the quasi-energy spectrum via the free fermion solution. The result of examining the quasi-energy degeneracies $\Delta_{12}$ confirms the location of the phase boundary obtained via the winding number, in which the topological phase is characterized by $\Delta_{12}=1$ with the value $\Delta_{12}=0$ observed in the trivial phase. Here $\Delta_{12}=0$ refers to the two smallest quasi-energies which occur as a pair of the same magnitude, which is not supposed to occur in Hermitian systems. 
For NH Hamiltonian $H_1$ the ring of EPs identified from the quasi-energy condition $\Delta_{12}=0$ 
coincides with the phase boundary $h_{R}^2+h_{I}^2=1$ in the infinite size limit~\cite{10.21468/SciPostPhys.15.1.016}. 

In contrast
the systems will have one unique quasi-energy satisfying $\epsilon=0$ when the system has $\Delta_{12}=1$. In other words, the existence of zero quasi-energy $\epsilon=0$ indicates the topological invariant. Although quasi-energy degeneracy characterizes the phase diagram successfully, it fails to differentiate the two topological phases with $w = \pm 1$. In this way we see that the quasi-energy degeneracy not only provides  information on EPs but also provides insight into NH topological properties. In the previous theoretical predictions for NH systems, the real energy gap should carry a pair of Majorana edge modes in the topological phase and an imaginary energy gap with no Majorana zero modes should exist in the trivial phase~\cite{Zhou_2021}.

It should be stressed that the exact solution of the NH free fermion models have only been discussed here for open boundary conditions, and the solution does not appear to guarantee the calculation of eigenvectors and order parameters. In the following part of the results, we adopt Exact Diagonalization (ED) to provide insight into critical phenomena associated with NH phase transitions the models under consideration. We report numerical results obtained from the biorthogonal fidelity, biorthogonal entanglement entropy, and dynamic behavior to study the groundstate properties of associated phase transitions. In the following, we identify the ground state as the state with the lowest real part of the energy and employ PBCs.

\begin{figure}
\centerline{\includegraphics[width=1.2\linewidth]{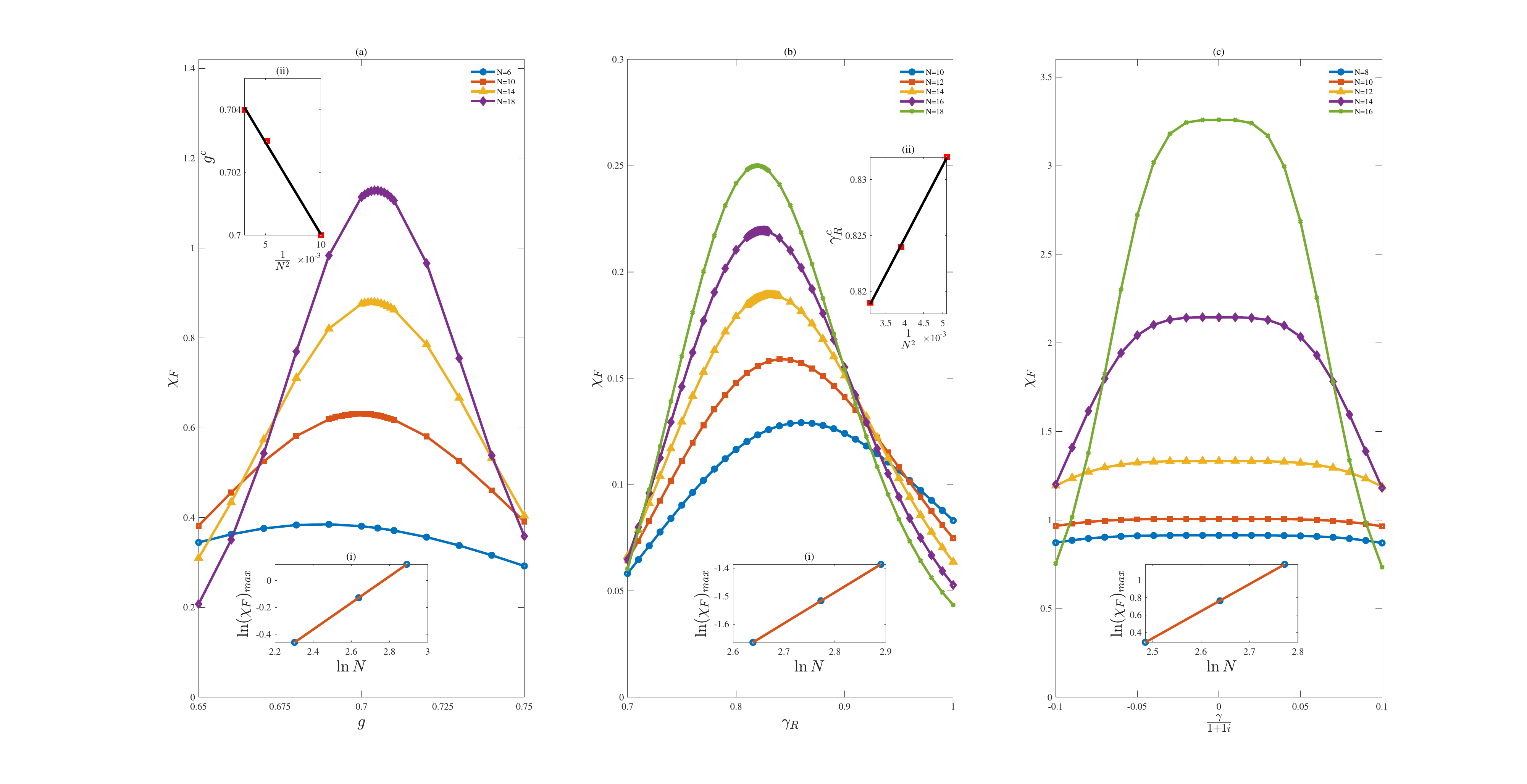}}
\caption{The scaling behavior of the fidelity susceptibility $\chi_F$ for the NH extensions of the XYh model with respect to different parameters and increasing system size $N$. (a) The Ising model with complex field with respect to $g=h_R/h_I$, (b) The XY model with purely imaginary field at $h_I=0.8$ with respect to $\gamma_R$, (c) The NH XY model with real field at $h_R=0.3$ with respect to $\gamma$. For each case the inserts (i) show the finite-size scaling of the maxima of $\chi_F$ and (ii) show the finite-size scaling of the critical points at the maxima of $\chi_F$.} 
\label{Fig2}
\end{figure}

We start our numerical trials with Hamiltonian $H_1$, the 
NH Ising model with complex field. Without loss of generality, we first study the properties of phase transitions where both the real field and imaginary field exist, by setting $g=h_{R}/h_{I}$. In this case, the critical point $g_c$ should be located at $g_c={\sqrt{2}}/{2}$ from the above analysis. From our results in Fig.~\ref{Fig2}(a), the peak of $\chi_F$ increases with the system size and tends towards the exact position $g^c={\sqrt{2}}/{2}$. The critical exponent $\nu$ from the finite-size scaling theory is shown in the insert of Fig.~\ref{Fig2}(d), where linear fitting yields $\nu=1.0060$. Meanwhile, we obtain the critical point $g^c=0.7069$ in the thermodynamic limit for $H_1$ by extrapolating data with $g^c(N)=g^c-a/N^2$, consistent with theoretical expectation. One should be aware that the Hamiltonian $H_1$ contains two special cases, by setting $h_I=0$ and $h_R=0$.  While the former case is reduced to the traditional transverse Ising model, the latter case is considered to be included in the Hamiltonian $H_2$. When the system interacts with the purely imaginary field, there is a PT transition occuring on the first and second excited states by applying the similarity transformation~\cite{PhysRevB.101.245152,PhysRevB.110.014441}. We consider the effect of a purely imaginary field by taking $h_I=0.8$ for model $H_2$ for which the corresponding phase transition point is located at $\gamma_{R}^c=0.8$, while the critical point obtained from extrapolation of finite systems sizes systems is at $\gamma_{R}^c=0.7988$. As seen from Fig.~\ref{Fig2}(b), the critical exponent $\nu$ describing the phase boundary line in $H_2$ from the finite-size scaling theory suggests $\nu \simeq 0.9524$. The phase transitions between the topological phase and the trivial phase in both $H_1$ and $H_2$ are separated by the critical line with critical exponent $\nu \simeq 1$, belonging to the same universality class and have the Hermitian counterpart, i.e., the Hermitian Ising class. We now turn to the phase transition between the two topological phases occurring for Hamiltonian $H_3$. Fig.~\ref{Fig2}(c) illustrates the behavior of $\chi_F$ as a function of $\gamma$ at $h_R=0.3$, with the corresponding scaling analysis shown in the insert of Fig.~\ref{Fig2}(c). The critical exponent $\nu \simeq 0.4870$ fails to follow the anisotropic phase transition in the Hermitian case but matches with the analytical derivation in a previous study with $\nu \simeq 1/2$~\cite{PhysRevX.4.041001}.

\begin{figure}
\centerline{\includegraphics[width=1.2\linewidth]{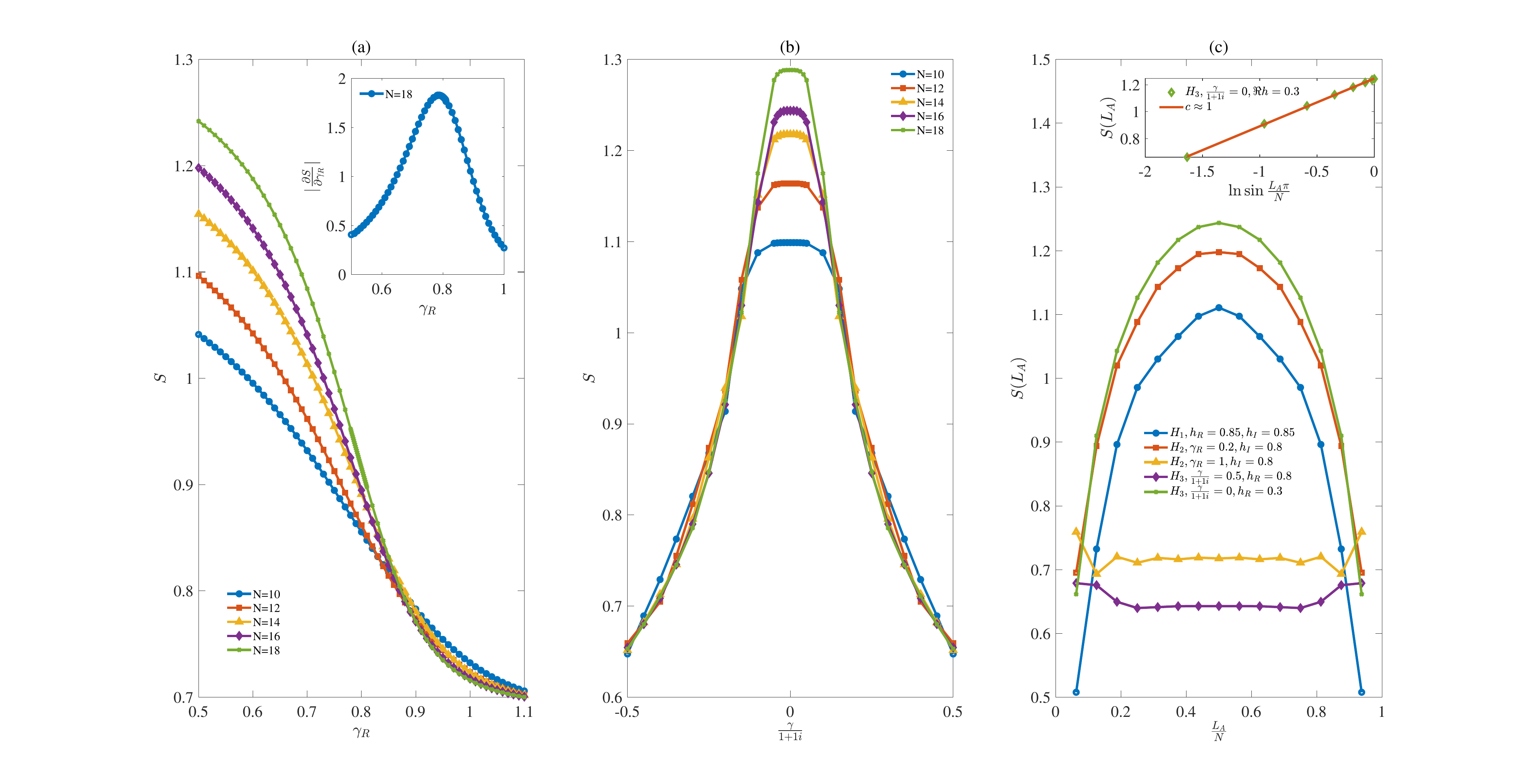}}
\caption{The biorthogonal entropy $S$ of the NH XYh model with respect to different parameters and system sizes. (a) Entropy $S$ of the NH XYh model with complex field at $h_I=0.8$ as a function of $\gamma_R$ in which the topological-trivial phase transition occurs, (b) Entropy $S$ of NH XYh model with real field at $h_R=0.3$ and varying $\gamma$ in which the topological-topological phase transition occurs, and (c) subsystem dependence of $S$ for fixed total system size $N=16$.} 
\label{Fig3}
\end{figure}

Next we examine the behavior of the biorthogonal entropy to describe the two different phase transitions occuring in Hamiltonians $H_2$ and $H_3$. We first calculate the half-chain bipartite entanglement entropy $S$ by choosing the subsystems $A$ and $B$ with same size, i.e., $L_A={N}/{2}$. Fig.~3 plots $S$ as the function of driving parameters $\gamma_R$ and $\gamma$ in $H_2$ and $H_3$. For the Hermitian phase transition occuring on other excited states than the ground state, like the BKT phase transition, the derivative of the entropy may be able to detect the location of the critical point~\cite{PhysRevB.104.205112}. We plot the first derivative of the entropy $\frac{\partial S}{\partial \gamma_R}$ in the insert of Fig.~\ref{Fig3}(a). The $\frac{\partial S}{\partial \gamma_R}$ values successfully pin out the critical point by its diverging behavior around the critical point $\gamma_R^c=0.8$.  Accompanying the topological-trivial phase transition, an entanglement phase transition is recognized via the different growth behavior of $S$ in Fig.~\ref{Fig3}(a). In the trivial phase ($\gamma_R <0.8$), the entropy $S$ does grow with increasing system size and shows the log-law behavior which is similar to that of the critical Hermitian model, while $S$ remains stable with increasing system size in the topological phase ($\gamma_R > 0.8$). Additionally, the behavior of $S$ witnesses a phase transition between two distinct topological phases, with $S$ having its maximum value at critical point $\gamma=0$. Interestingly, the logarithmic behavior emerges at the critical point between these two area-law phases. This is similar to what occured in the NH Aubry-Andr\'e-Harper model~\cite{PhysRevB.109.024306}. The behavior of the entropy $S$ in the topological-topological phase transition satisfies the volume law at its critical point and exhibits area law in the adjacent topological phase. 
Fig.~\ref{Fig3}(c) shows the relationship between entropy $S(L_A)$ and $L_A$ for fixed $N=16$ at different parameters for Hamiltonian $H_1$, $H_2$ and $H_3$. In the topological phase ($\{ \gamma_R=1,h_I=0.8\}$ in $H_2$ and $\{ \frac{\gamma}{1+1\mathrm{i}}=0.5,h_R=0.3\}$ in $H_3$), the values of $S(L_A)$ almost do not vary with $L_A$, scaling as the area-law $S(L_A)\sim L_A^{d-1}$, while the values of $S(L_A)$ in the trivial phase ($\{ h_R=0.85,h_I=0.85\}$ in $H_1$ and $\{ \gamma_R=0.2,h_I=0.8\}$ in $H_2$) are described by the fitting function $a \ln [\sin (\pi L_A /N)]+b$. Also, within the prediction of CFT theory, the central charge at critical points ($\{ \frac{\gamma}{1+1i}=0,h_R=0.3\}$ in $H_3$) can be fitted through Eq.~(\ref{eq21}), from which numerical fitting yields the central charge $c \simeq 1.065$. This value is reasonably close to the exact value $c=1$ for the anisotropic phase transition occuring in the Hermitian XY model. Although it seems that the NH anisotropic phase transition is also characterized by the central charge $c=1$, the non-Hermiticity does affect how the system approaches the critical points and yields an unusual critical exponent $\nu$.

\begin{figure*}
\centerline{\includegraphics[width=1.2\linewidth]{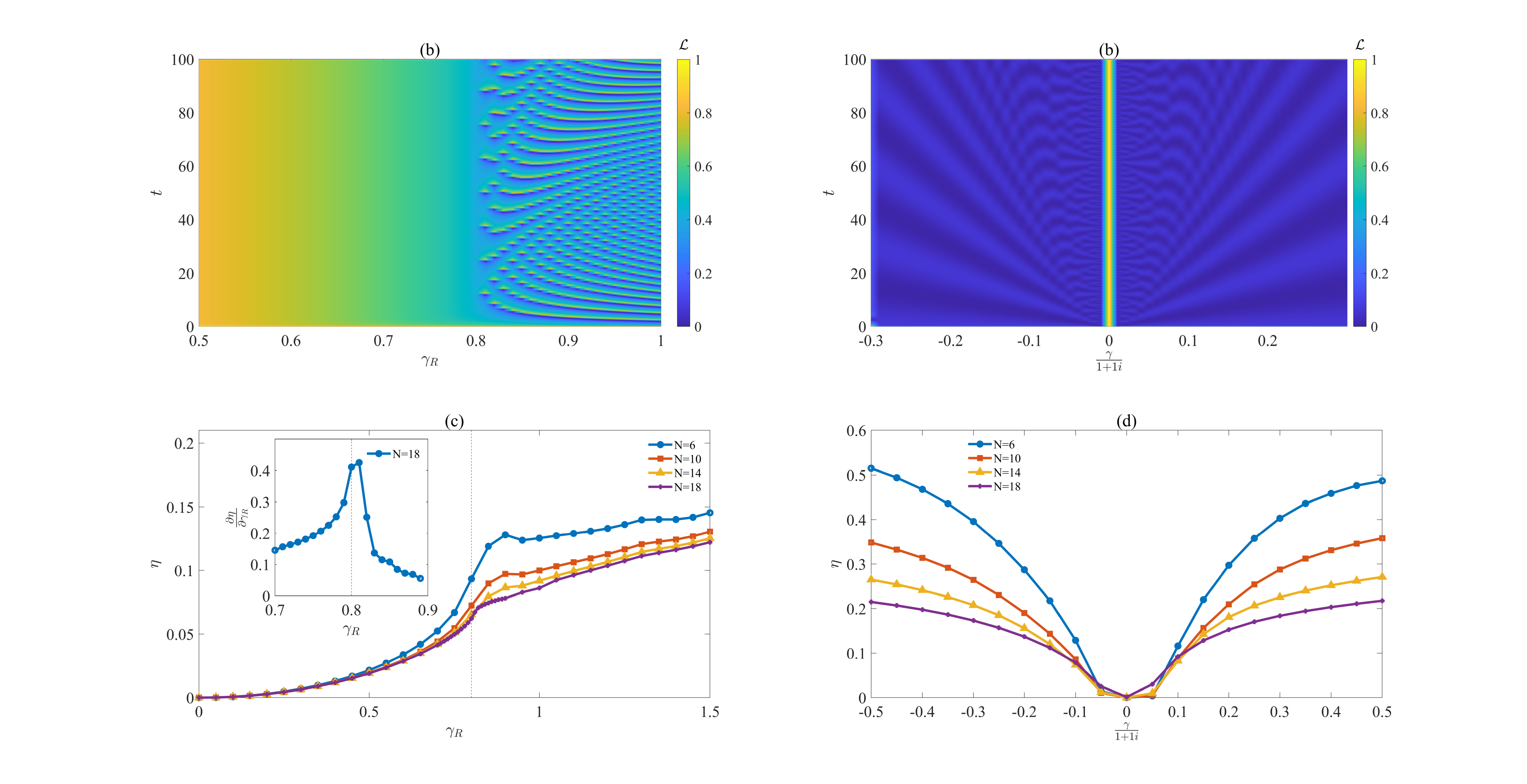}}
\caption{The dynamic behavior of the Loschmidt echo $\mathcal{L}$ with respect to different parameters for $N=10$ for (a) topological-trivial phase transition in Hamiltonian $H_2$ at $h_I=0.8$ and (b) topological-topological phase transition in Hamiltonian $H_3$ at $h_R=0.3$. (c) and (d) The time average behavior of Loschmidt echo $\eta$ as a function of driving parameter corresponding to (a) and (b) for different system sizes.} 
\label{Fig4}
\end{figure*}

Within the framework of NH physics, systems will evolve into a steady state with the largest imaginary energy. Here through the dynamics of the Loschmidt echo $\mathcal{L}$ demonstrated
in Fig.~\ref{Fig4}(a), the transition between the topological phase and the trivial phase becomes evident, specifically around the critical point $\gamma_R^c = 0.8$. In
detail, $\mathcal{L}$ shifts from monotonic behavior (trivial phase) to oscillatory behavior (topological phase). While the topological phase witnesses the oscillation characteristics of $\mathcal{L}$, the time decay of $\mathcal{L}$ with  stabilization in the long time limit is evident in the trivial regime. We are also interested in the dynamic behavior of $\mathcal{L}$ associated with the topological-topological phase transition, as depicted in Fig.~\ref{Fig4}(b). A notable change in the behavior of $\mathcal{L}$ is observed at the critical point $\gamma^c=0$, clearly indicating a phase transition at $\gamma=0$ while the existence of oscillation of $\mathcal{L}$ in the topological phase is not as pronounced as in Fig.~\ref{Fig4}(a). Our numerics show the one-to-one correspondence relation between the equilibrium topological phase diagram and DQPTs. Here the DQPTs are only found when the system is quenched to a NH topological phase. To provide a more deep connection between dynamics and NH phase transitions, we introduced the time-average of the Loschmidt echo $\eta$ by calculating $\eta= -\frac{1}{N} \ln [ \lim_{t \to \infty} \frac{1}{t}\int_0^{t} \mathcal{L} dt]$, which is also called time average of rate function. The values of $\eta$ will exhibit non-analytic behavior when the system approaches the critical point and therefore can distinguish between phases. This can be seen in  Fig.~\ref{Fig4}(c) and (d), where $\eta$ is plotted as a function of driving parameters for different system sizes. The non-distinct kink behavior of $\eta$ is evident as the phase transition points $\gamma_R^c=0.8$ and $\gamma^c=0$ are crossed in figures (c) and (d) respectively. The phase transition point is further highlighted by the discontinuous behavior evident in the first derivative, as shown in the insert of Fig.~\ref{Fig4}(c). All of the numerical results are self-consistent and give us a deeper understanding of criticality in NH many-body systems.

\section{Conclusion}

In summary, we have examined NH phase transitions and  critical behavior in NH extensions of the XYh model.
 The non-Hermiticity in the models under consideration is not only generated from single-site loss but also arises from the anisotropic coupling. By making use of the general open boundary free-fermion solution and the theory of NH topological invariants, we depicted the model phase diagram by considering three extensions of the non-Hermitian XYh model: the Ising model with complex field, the XY model with  complex field, and the NH XY model with real field. The change of the NH winding number witnesses the NH phase transition between the topological phase and trivial phase and the NH phase transition between the two topological phases. By analyzing the structure of the quasi-energy spectrum, we found that the NH topological phase has one unique quasi-energy satisfying $\epsilon=0$. The NH trivial phase is characterized by the degeneracy of quasi-energies. Meanwhile, in our model, the NH topological-trivial phase transition is associated with the entanglement phase transition. The biorthogonal half-chain entropy in the NH topological phase exhibits the area-law behavior and shows log-law scaling in the trivial phase, which is similar to Hermitian critical spin models. 
 The critical boundary of the topological-topological phase transition exhibits the characteristics of
logarithmic relation and the central charge $c \simeq 1$ is fitted through subsystem finite size scaling. Additionally, from the behavior of the fidelity susceptibility and its scaling analysis, the critical behavior in the different cases shows that the NH models can maintain the universality class of the Hermitian models, but is also able to have an unusual value of the critical exponent $\nu$. The results for the Loschmidt echo and its time average behavior confirm the theoretical prediction that the dynamic phase transition will only be observed when the system is quenched to nontrivial NH topological phases.

Although they can be viewed as relatively simple in the realm of exactly solved models, as we have seen here in the context of the paradigmatic XY model in a magnetic field, free fermion models have interesting and worthwhile NH extensions, which shed further light on the study of NH physics in many-body systems. In particular, there is an established intrinsic connection between free fermion quasi-energies and topological invariants.

\begin{acknowledgments}
D.C.L thanks Xilin Lu and Zixiang Li for helpful discussions.   
This work has been supported by
Australian Research Council Grants DP210102243 and DP240100838.
\end{acknowledgments}

\appendix

\section{Energy spectrum in small systems}

\begin{figure*}[htb]
\centerline{\includegraphics[width=1.0\linewidth]{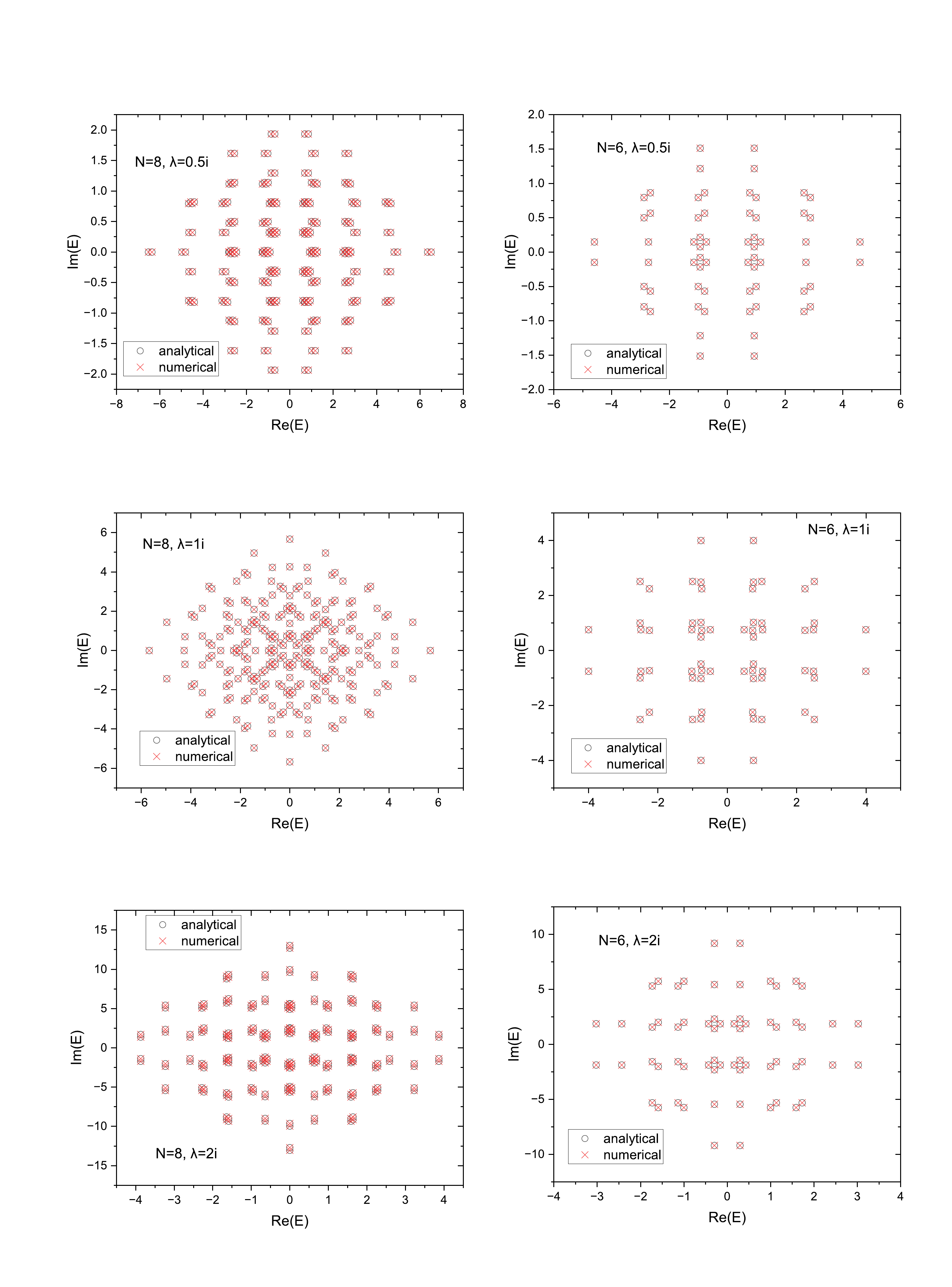}}
\vskip -10mm
\caption{Complex energy spectra for the NH extension of the XY model with various values of system size $N$ and coupling $\lambda$ (see text). The energy  values obtained from the quasi-energies (black circles) are compared with the values obtained from the exact diagonalization of the Hamiltonian (red crosses). In all cases, there is perfect agreement.} 
\label{Fig5}
\end{figure*}

In general it can be a challenge to study NH systems for large system sizes, as doing so requires a higher numerical cost than their Hermitian counterparts. In this Appendix, we illustrate in Fig.~\ref{Fig5} the use of numerical full exact diagonalization to examine the correctness of the free fermion energy spectrum given in Eq.~(\ref{eq8}) for a range of different complex parameters. Here we have rescaled the Hamiltonian in Eq.~(\ref{eq1}) with $\lambda=\frac{1-\gamma}{1+\gamma}$ and $h=0$. In addition to showing the overall structure of the complex energy spectrum, the results demonstrate the validity of the exact results. Since the energy spectrum is known exactly via the free-fermion solution for large system sizes, it provides a useful benchmark for numerics to  test that the non-Hermitian eigensolver works correctly in other algorithms like DMRG.

\bibliography{reference}
\end{document}